\documentclass[a4paper,fleqn,usenatbib]{mnras}

\usepackage{txfonts}
\usepackage[T1]{fontenc}
\usepackage{ae,aecompl}

\usepackage[multiple]{footmisc}
\usepackage{graphicx}

\title[surveying PIs with unpublished data]{Closing the loop: \\surveying PIs who have not published their data}

\author[F. Stoehr et al.]{
Felix Stoehr,$^{1}$\thanks{E-mail: fstoehr at eso dot org}
Erik Muller,$^2$
Mark~Lacy$^3$
and St{\'e}phane~Leon~Tanne$^4$
\\
$^{1}$ESO/ALMA, Karl-Schwarzschild-Str. 2, 85748 Garching, Germany\\
$^2$NAOJ, 2-21-1 Osawa, Mitaka, Tokyo 181-8588, Japan\\
$^3$NRAO, 520 Edgemont Road, Charlottesville, VA 22903-2475, USA\\
$^4$JAO, Alonso de Cordova 3107, Vitacura - Santiago, Chile
}

\date{}

\pubyear{2016}

\begin{document}
\label{firstpage}
\pagerange{\pageref{firstpage}--\pageref{lastpage}}
\maketitle

\begin{abstract}
With high over-subscription rates and significant operational costs, observatories must ensure that their operations are efficient and effective.  A number of key performance indicators are generally used to evaluate the observatory's performance among which are the numbers of publications and citations of refereed journal articles to measure the overall scientific impact. Those measures, however, are broad and can not assess whether the observatory was successful on a project-by-project basis to deliver data to the PIs enabling them to carry out their science and to publish their results. In particular the reasons that prevented PIs from publishing remain hidden. Understanding and acting upon those reasons, however, have the potential to substantially improve the observatory's operational model. Of course not every approved project even should lead to a publication. Indeed, the risk of not finding the expected (or any unexpected) science in the data the PI receives is an inherent and indispensable part of the scientific process. But even here, measuring the fraction of such projects can lead to valuable insights which might then be used to instruct future proposal review committees. To fully close the loop on the end-to-end data-flow, ALMA has started in March 2015 to send survey questions to PIs where two years after the end of the proprietary period no publication making use of the delivered data could be identified. We describe our method as well as the type of conclusions we hope to be able to draw once a statistically relevant sample of answers has been received.
\end{abstract}


\section{Introduction}
The major astronomical facilities typically receive hundreds, some more than thousand observing proposals for each of the half-yearly or yearly Calls for Proposals they issue. The requested time typically exceeds the available observing time by factors of 2 to 9. \footnote{http://almascience.org/documents-and-tools/cycle4/c04-proposal-review-process}\footnote{http://www.eso.org/sci/observing/phase1/p99/pressure.html}\footnote{http://www.stsci.edu/institute/stuc/april11/whitmore.pdf}\footnote{http://www.noao.edu/gateway/tac/obsreqs16a\_s.html}\footnote{http://irsa.ipac.caltech.edu/data/SPITZER/docs/files/spitzer/go13-stats.pdf}

Subsequently the facilities deploy large efforts to select the scientifically highest ranked of the submitted proposals until the available observing time is distributed. In general, teams of up to 150 international experts meet for several days face-to-face to balance the scientific merit of the proposals against each other and to produce the final ranked list. 

Such an expensive and careful process is mandated by the extremely valuable observing time. Indeed, by just dividing the annual total budget of an observatory by the number of the observatory's refereed publications over a year we obtain a back-of-the-envelope estimate of the cost of a scientific publication. This estimate of course is very crude and ignores the construction budget but also ignores the added value of staff-science, of R\&D, training of young scientists, outreach etc. Nevertheless for our purposes here this is good enough and we find that for many facilities this value is in excess of 50kEUR/publication.

As a consequence, since long, observatories are very carefully monitoring the success of their scientific program and are trying to improve their operational model. Most major facilities are tracking the publications that made use of the data taken by the facility sometimes together with statistics of the number of citations (recently e.g. \citet{2010SPIE.7737E..1VS}, \citet{2010PASP..122..808A}, \citet{2012PASP..124..391R},  \citet{2014AN....335..210N}, \citet{2014SPIE.9149E..0AC}, \citet{2015salt.confE..72S}, \citet{2015ASPC..492..133S}, \citet{2015Msngr.162...30S}). 

In addition, most large observatories are running user-surveys with their communities. These typically ask questions concerning the entire end-to-end data-flow with the aim of improving the overall operations and user-experience. 

We expect the need of the facilities to measure their scientific impact to grow with time. This is not only driven by the funding agencies but especially by the growing amount of data with time. A full SKA alone is expected to deliver up to 10PB/day of science images. This corresponds to about 180TB/year of images (or twice the entire data output of ALMA in 2016 or alternatively 50 million 1kx1k FITS images) for each and every astronomer registered with the IAU at that time. With most pixels that get observed never being looked at by a human, it is indeed possible that only a fraction of the data taken get also analysed and published.

We also expect that the need of the facilities to measure their effectiveness and efficiency will grow because over the decades they have and will be taking up more and more of the responsibility of the scientific process. Already the observations themselves (service-mode observing) and the data reduction (data-reduction pipelines of HST, ESO, XMM, SPITZER, ALMA, VLA, etc.) and sometimes even the first science analysis (ALMA Regional Centres, ADMIT \citep{2015ASPC..495..305T}) are already taken up by the observatories (see also  \citet{2014SPIE9149E02S}). 

While these publication statistics and the general user-surveys measure the overall success and user-experience of a facility as a whole, they hide the issues PIs may encounter on an individual project level. For example it is entirely conceivable that the overall publication statistics show a quite large number of publications, but still only a relatively small fraction of projects of each proposal cycle actually lead to a publication (e.g. \citet{2015Msngr.162....2S}).

\section{Motivation}
We propose here therefore to fully close the loop of the end-to-end data-flow by identifying the projects that have not led to a publication a certain time after the end of the proprietary period of the data of the project, and then directly contacting the PIs of those projects individually to find out the underlying reasons. 

Such a survey is expected to be able to uncover stumbling blocks and provide guidance for improved operations that would be hidden in the noise of a general user-survey, would be incomplete or would not even appear there at all, due to the fact that PIs who have not published may not be part of the 10\% of users who typically do fill out such general surveys in the first place. 

Such a dedicated survey also goes beyond the request to the PIs in proposal submission tools to indicate how they have used their previously delivered data. In particular, a survey of PIs with unpublished data allows for a statistical analysis and -- more importantly -- directly asks for the reasons for non-publication.

We note that effective time-allocation is not only in the best interest of the observatories and the funding agencies but also of the PIs themselves, as PIs who do not publish their data implicitly are also penalizing PIs who were unsucessful in obtaining telescope time, but who otherwise would have carried out a successful scientific program. 

Identifying the unpublished projects is possible for telescopes like the VLT, ALMA or HST, for which the librarians not only track the publications themselves but also link the publications to the data that was used in the publication. Such linking is enabled by setting up policies that require authors to put the dataset IDs of the data they have been using into their publication \citep{2014SPIE.9149E..26M}.

The reminder of this paper is organised as follows. We first highlight some general considerations in section \ref{section:considerations} before we briefly describe our method in section \ref{section:method}. We then give an overview about the current status in section \ref{section:results} and conclude in section \ref{section:conclusions}.

\section{considerations}
\label{section:considerations}
Asking the PIs of successful proposals why they did not publish their data is delicate. Certainly, in some cases, an embarrassment of the PI can not be fully excluded. Also, with rapidly developing scientific fields and many astronomers submitting proposals to a variety of facilities, inevitably the data of some projects gets less attention of analysis by the PI than others. This leads to a situation where after some time essentially every observational astronomer will end up with some of the data they have received that they did not analyse and publish in a refereed journal. This is the everyone-has-a-skeleton-in-the-closet problem. 

The main consideration therefore was to carefully formulate the text and questions to the PIs so, that it is unambiguously clear that the goal of the survey is not to embarrass PIs, but to {\it learn} about the facility's deficiencies and to be able to improve upon them. The email that gets send to the PIs as well as the text on the survey page itself (\ref{fig:survey}) have therefore been crafted meticulously to avoid the impression of harassment as much as possible.

Given the high operational cost to take the data for a proposal, however, management considered it acceptable for the PIs in return for the data they receive and the proprietary period they get assigned to also answer such a survey. These views were also confirmed by the ALMA Science Advisory bodies (ESAC, ASAC).

In addition, it was also decided by ALMA management that the survey should be semi-anonymous. While a token should be issued for each unpublished program, and while it should be recorded whether or not a given PI has already responded, the answers were to be kept separately and without link to the tokens so that answers and PIs can not be related any more. While it is clear that the analysis can then only be global and not related to the type of project, the wish to protect the PIs as much as possible prevailed.

A further important consideration was to make clear to everyone involved -- from the PIs to the management who will receive the compiled statistics --, that of course by no means it is expected that each PI project that receives data from the facility should result in a publication. The contrary is the case. Non-detections or observations where neither the expected science nor any unexpected science are contained in the data are a natural and indispensable part of the scientific discovery process. Observatories are longing for science with very high impact (high-reward science) which often enough comes with high-risk proposals. It is one of the most beautiful properties of science that it can hardly be planned (\citet{cosmic}, \citet{1611.05570}).

That said, knowing the fraction of projects that fall into category of unpublished data because they were high-risk/high-reward is certainly a very important indicator for a facility. Indeed, the risk that that due to very high oversubscription factors and a huge step of improved capabilities of a new facility like ALMA, the Proposal Review Committee (PRC) in the first years may be rather conservative is real. Should  such a result be found during the survey, PRC members could be instructed to rate high-risk/high-reward projects higher.

The final consideration was that the survey should be running continuously to allow the observatory to first adapt to the results as early as possible but then also to evaluate as soon as possible the impact of the changes that have been applied to the operational model. By setting a fixed time-span after which PIs would be asked to provide feedback, we also remove time-span-related biases.

\section{Method}
\label{section:method}
Publications and their links to the corresponding science data in the ESO Archives, including ALMA, are tracked by the ESO Library in collaboration with their colleagues from NRAO and NAOJ and then stored in a database (e.g. \citet{2010ASPC..433...81E}, \citet{2012ASPC..461..767M}, \citet{ 2015ASPC..492...63G}). Information about the data deliveries, delivery dates and proprietary periods is contained in the ALMA Science Archive database. 

We have developed software that runs daily and identifies projects which do not have a related publication two years after the end of the proprietary period of the last data delivery to the PI. The PIs of any such projects then automatically get sent the personalised email containing the link and token for the survey called "project status questionnaire" which is hosted on the ALMA Science Portal. Should there be also no publication three months later and should the PI have not yet filled out the survey, a reminder email is sent, explaining again the high value the answers do have for the project. Should a European ALMA PI not have answered after the reminder email, the the policy allows for this PI be called by phone. So far, such PIs have been extremely polite and understanding.

The choice of the time-span of two years after the end of the proprietary period was motivated by two factors. On the one hand, the more timely the information from the PIs is received, the faster the observatory can react. On the other hand, in order to maximise the usefulness of the survey, the time-span should be long enough that enough time was available to produce a publication under normal circumstances. 

\begin{figure}
	\includegraphics[width=\columnwidth]{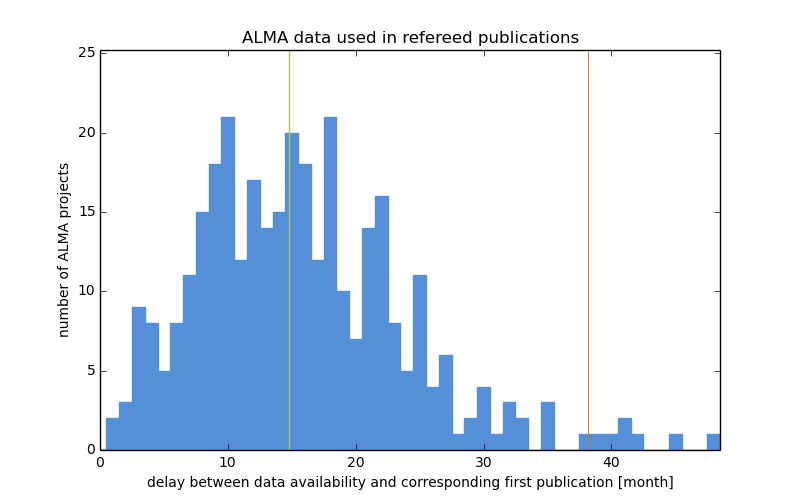}
    \caption{Publication delay. Distribution of the time-span between the delivery of data to the PI and the first refereed publication making use of those data. In case the PI has received several data deliveries for his project, the median delivery date is used as start date. The green line indicates the median publication delay of 14.8 months, the orange line indicates the time at which ALMA seeks feedback from the PIs of unpublished projects. }
    \label{fig:publicationdelay}
\end{figure}

Fig. \ref{fig:publicationdelay} shows the distribution of publication delays, i.e. the distribution of the time-spans between the (median) delivery time of data to the ALMA PI and the moment of the first related publication. The median value of the publication delay is 14.8 months as indicated by the green line. These publication delays are very short compared to other facilities (see also \citet{2015Msngr.162...30S}). Therefore, using a time-span of two years after the end of the proprietary period, as indicated by the orange line, is indeed a safe choice for sending out the survey email. 

To keep the barrier as low as possible, the survey was designed to fit onto a single page (see Fig. \ref{fig:survey}). On top of the page, again, the text from the email is repeated to avoid all potential ambiguity as to what the aim of the survey is. A single comment box allows PIs to elaborate on their answers, but only the selection of the reason for non-publication is mandatory in this survey. 

The survey questions are presented to the PI in roughly ascending order of degree of potential embarrassment:
\begin{itemize}
\setlength\itemsep{1em}
\item {\it There is a publication}. We hope to learn from answers to this question whether or not the set of journals we currently monitor (A\&A, A\&ARv, AJ, AN, ApJ, ApJS, ARA\&A, EM\&P, Icarus, MNRAS, Nature, NewA, NewAR, PASJ, PASP, P\&SS, Science) is complete. 

\item {\it A publication is in press} and  {\it a publication is in preparation}. Some projects take long time to analyse, especially when data from different facilities are combined. 

\item {\it Only part of the requested project was actually observed}. ALMA is working hard on increasing the fraction of projects that are fully completed, but in Cycles 1 and 2 there were still about 25\% of the high-priority projects which were only partially completed. While we do know the completion fraction, we want to learn what impact the delivery of non-completed projects does have on the PIs possibility to publish. The findings can then directly be used to optimize the scheduling algorithm which has to balance the completion of already started but unfinished projects agains the observation of new projects.

\item {\it The quality of the data was not good enough}. The User Survey shows that generally the quality of the ALMA data is very high. If however, in some cases the quality of the data is the single most important reason preventing the PI from being able to write a publication, the observatory would need to give a high priority to the investigation of the issues and to improving the data-quality to avoid the waste of observing time.

\item {\it We had problems producing the data products}. Calibration and Imaging are complex tasks in radio interferometry. To this end ALMA is developing the CASA software package \citep{2010AAS...21547904R} which now includes the ALMA Pipeline (\citet{2015ASPC..499..355S} an references therein). Should this reason be the show-stopper for a non-negligible fraction of projects, the tools but also the help provided to the PIs would need improvement.

\item {\it The expected science was not contained in the data}. As mentioned earlier, we try to evaluate the fraction of high-risk/high-reward programs allocated by the PRC. The findings here can directly be fed back into the proposal evaluation process.

\item {\it The scientific field had moved on in the meantime}. With yearly proposal cycles, ALMA's turn-around time is relatively long compared to some other facilities (e.g. the VLT). This question helps us evaluate how much this fact impacts science. Should this turn out to be a significant factor, then two cycles per year could be offered or, more drastically, a fast-turnaround queue like for Gemini \citep{1408.5916} could be implemented. 

\item {\it No effort was available any more}. The times from proposal submission to data delivery can be long enough to have an impact of the availability of scientists, e.g. PhD students in a research group. As above, a large fraction of responses selecting this option would suggest to reduce the turn-around time of the end-to-end data-flow.

\item {\it Waiting for other facilities}. With astronomy transforming to becoming fully multi-wavelength science, the needs for coordinated proposal processes of different facilities might grow. We hope to evaluate this need and to find out which other facilities are mostly required by the PIs. Possible subsequent actions could be to implement common joint observing programs like they do exist for the VLT and XMM.

\item {\it Personal reasons} and {\it Other}. These give the PIs the possibility to complete the survey without having to provide substantial information or to bring up a reason we have not thought of.
\end{itemize}

As mentioned, the survey was set up to run continuously right from the start of ALMA operations, i.e. right from the point in time when the first projects could fulfil the two-year criterion.

\section{Preliminary results}
\label{section:results}
The first survey email was sent out in March 2015 and the software has been running and the survey has been ongoing since. As it turns out, the publication fraction, i.e. the fraction of all ALMA projects that do receive a publication, is extremely high \citep{2015Msngr.162...30S}. Currently that fraction is hovering around 90\% (including archival publications). While this is a very large success for ALMA, it also means that the number of answers to the survey is low. 

To this date, 21 emails to PIs have been sent out and a total of 8 answers have been received. Four PIs have been called on the phone. While no solid statistics can be presented at this stage, the answer "A publication is in preparation" is so far the one given most often, followed by "Only part of the requested project was actually observed" and then at the same level "There is a publication", "A publication is in press" and "We had problems producing the data products". Out of the 8 responses, however, 7 PIs provided comments and all of those comments are extremely useful. Given the very targeted survey, the return-rate is below the expectations at this stage. 

We are planning improvements for this survey in the medium-term. These potentially could include sending out emails even if an archival publication has been made by other users before, converting of the radio-buttons into check-boxes allowing PIs to select several reasons at the same time, updating the text in the email to insist more on the fact that the survey is short, adding to the ALMA User's Policy that by submitting a proposal to ALMA PIs are agreeing to fill out the survey, explaining in the email that the statistics will be presented to ALMA management at all levels, asking the Contact Scientist of each project to follow up with PIs who have not filled out the survey, changing the survey to be non-anonymous (while at the same time of course keeping the individual answers confidential) to be able to link the reasons to the types of the project as well as sending the list of PIs who have not answered to management or the PRC.

\section{Conclusions}
\label{section:conclusions}
We have presented a method to close the loop on the end-to-end data-flow on a project-by-project level by asking PIs of programs which have received data but where no related publication could be identified after 2 years for feedback. Such a targeted survey has been implemented and is continuously running. To our knowledge, ALMA ist the first astronomical facility directly surveying PIs of unpublished data. 

While due to the high publication fraction of ALMA data the number of answers to our survey are too low to draw statistically relevant conclusions at this stage, the answers and the quality of the comments received indicate that surveying the PIs with unpublished data has a great potential to identify stumbling blocks and as a consequence to improve operational model of the facility. Also science advisory bodies are showing very high interest in the results of this survey project.

\section*{Acknowledgements}
We thank Kouichiro Nakanishi, Paola Andreani, Uta Grothkopf, Dominic Bordelon, Michael Sterzik, Martino Romaniello and Ferdinando Patat for very fruitful discussions. The presented project makes use of the ESO Telescope Bibliography (telbib), maintained by the ESO Library. ALMA publications are tracked in collaboration with colleagues from NRAO and NAOJ. ALMA is a partnership of ESO (representing its member states), NSF (USA) and NINS (Japan), together with NRC (Canada), NSC and ASIAA (Taiwan), and KASI (Republic of Korea), in cooperation with the Republic of Chile. The Joint ALMA Observatory is operated by ESO, AUI/NRAO and NAOJ.

\bibliographystyle{mnras}
\bibliography{unpublished}

\appendix

\section{Survey questions}

\begin{figure*}
	\includegraphics[width=\textwidth]{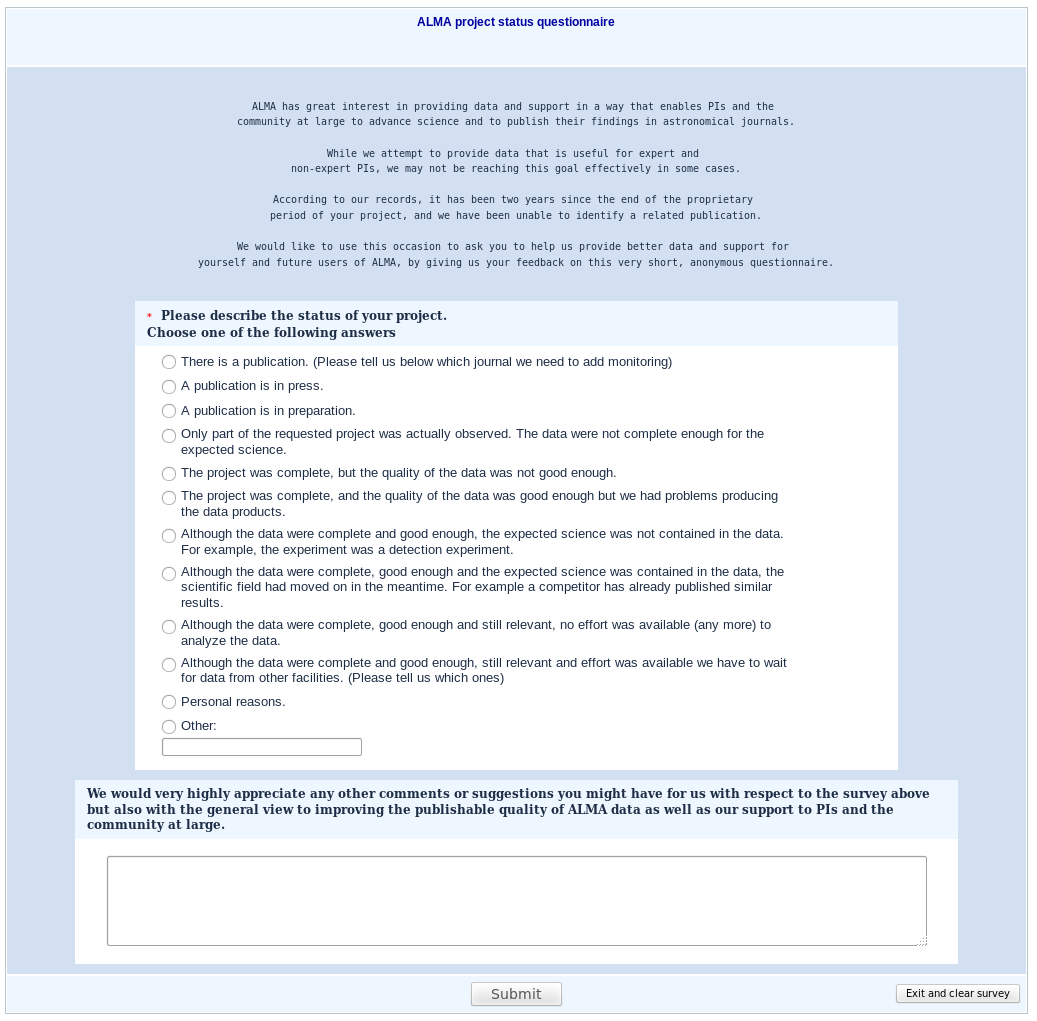}
    \caption{Survey form sent per personalised email to PIs two years after the end of the proprietary period of the last dataset from their project.}
    \label{fig:survey}
\end{figure*}

\bsp
\label{lastpage}
\end{document}